\documentclass{svproc}

\usepackage{amsmath}
\usepackage{amsfonts}
\usepackage{hyperref}
\usepackage{url}
\usepackage{graphicx}
\usepackage{multirow}
\usepackage{tikz}
\usetikzlibrary{arrows.meta}
\usepackage{subcaption}
\usepackage{wrapfig}

\newcommand{\reffig}[1]{Fig.~\ref{fig:#1}}

\newcommand{\reftbl}[1]{Table~\ref{tbl:#1}}

\newcommand{\lblfig}[1]{\label{fig:#1}}
\newcommand{\lblsec}[1]{\label{sec:#1}}

\newcommand{\lbltbl}[1]{\label{tbl:#1}}

\def\R{{\mathbb R}}
\def\expm{\operatorname{expm}}
\def\vol{\operatorname{vol}}
\def\diag{\operatorname{diag}}
\def\Diag{\operatorname{Diag}}

\newcommand{\camready}[1]{#1}
\newcommand{\rotatev}[1]{\rotatebox[origin=l]{90}{#1}}
\renewcommand{\b}[1]{\textbf{#1}}

\usepackage{url}

\begin{document}
\mainmatter              % start of a contribution
\title{Dissecting graph measure performance for node clustering in LFR parameter space}
\titlerunning{Dissecting graph measure performance...}  % abbreviated title (for running head)
%                                     also used for the TOC unless
%                                     \toctitle is used
%
\author{Vladimir Ivashkin\inst{1} \and Pavel Chebotarev\inst{2}}
\authorrunning{Vladimir Ivashkin et al.} % abbreviated author list (for running head)
%
%%%% list of authors for the TOC (use if author list has to be modified)
\tocauthor{Vladimir Ivashkin, Pavel Chebotarev}
\institute{Moscow Institute of Physics and Technology, Moscow, Russia,\\
\email{vladimir.ivashkin@phystech.edu}
\and
Institute of Control Sciences of RAS, Moscow, Russia,\\
\email{pavel4e@gmail.com}}

\maketitle              % typeset the title of the contribution

\begin{abstract}
Graph measures that express closeness or distance between nodes can be employed for graph node\camready{s} clustering using metric clustering algorithms. There are numerous measures applicable to this task, and which one performs better is an open question.
We study the performance of~25 graph measures on generated graphs with different parameters.
While usually measure comparisons are limited to general measure ranking on a particular dataset, we aim to explore the performance of various measures depending on graph features. Using an LFR graph generator, we create a dataset of \camready{11780} graphs covering the whole LFR parameter space. For each graph, we assess the quality of clustering with $k$-means algorithm for each considered measure. Based on this, we determine the best measure for each area of the parameter space. We find that the parameter space consists of distinct zones where one particular measure is the best. We analyze the geometry of the resulting zones and describe it with simple criteria. Given particular graph parameters, this allows us to recommend a particular measure to use for clustering.
\keywords{graphs, graph measures, clustering, \camready{kernel k-means}}
\end{abstract}

\section{Introduction}
Graph nodes clustering is one of the central tasks in graph structure analysis. It provides a partition of nodes into disjoint clusters \camready{(communities)}, which are groups of nodes that are characterized by strong mutual connections or similar external connections. It can be of practical use for graphs representing real-life systems, such as social networks or industrial processes. Clustering allows to infer some information about the system: the nodes of the same cluster are highly similar, while the nodes of different clusters are dissimilar. The technique can be applied without any labeled data to extract important insights about a network.

There are different approaches to clustering, including ones based on modularity optimization~\cite{newman2004finding,blondel2008fast}, label propagation algorithm~\cite{raghavan2007near,barber2009detecting}, Markov cluster process~\cite{van2000graph,enright2002efficient}, and spectral clustering~\cite{von2007tutorial}. In this work, we use an approach based on choosing a closeness measure on a graph, which allows one to use any metric clustering algorithm (e.g.,~\cite{yen2009graph}).

The choice of the measure significantly affects the quality of clustering. Classical measures are the \textit{Shortest Path}~\cite{buckley1990distance} and the \textit{Commute Time}~\cite{gobel1974random} distances. The former is the minimum number of edges in a path between a given pair of nodes. The latter is the expected number of steps from one node to the other and back in a random walk on the graph. There is a number of other measures, including recent ones~(e.g.,\cite{estrada2017accounting,jacobsen2018generalized}); many of them are parametric.
Despite the fact that graph measures are compatible with any metric algorithm, in this paper, we restrict ourselves to the kernel $k$-means algorithm~(e.g.,~\cite{fouss2016algorithms}).

We base our research on a generated set of graphs. There are various algorithms to generate graphs with community structures. The well-known ones are the Stochastic Block Model~\cite{holland1983stochastic} and Lancichinetti--Fortunato--Radicchi benchmark~\cite{lancichinetti2008benchmark} (hereafter, LFR). The first one is an extension of the Erdős--Rényi model with different intra- and intercluster probabilities of edge creation. The second one involves power law distributions of node degrees and community sizes. There are other generation models, e.g., Naive Scale-free Clustering~\cite{pasta2017topology}. We choose the LFR model: although it misses some key properties of real graphs, like diameter or the clustering coefficient, this model has been proven to be effective in meta-learning~\cite{prokhorenkova2019using}.

There are many measure benchmarking studies considering node classification and clustering for both generated graphs and real-world datasets, including \cite{fouss2012experimental,sommer2016comparison,sommer2017modularity,avrachenkov2017kernels,ivashkin2016logarithmic,guex2018randomized,guex2019covariance,aynulin2019efficiency,aynulin2019impact,courtain2020randomized,leleux2020sparse}. Despite a large number of experimental results, \camready{an exact theory is} still a matter of the future. One of the most interesting theoretical results on graph measures is \cite{luxburg2010getting}, where some unattractive features of the Commute Time distance on large graphs were explained theoretically, and a reasonable amendment was proposed to fix the problem. %%This paper shows that such proofs are quite difficult. 
Beyond the complexity of such proofs, there is still very little empirical understanding of what effects need to be proven.
Our empirical work has two main differences from the previous ones. First, we consider a large number of graph measures, which for the first time gives a fairly complete picture. Second, unlike the previous studies aimed at revealing the global leaderboard, we are looking for the leading measures for each set of the LFR parameters.

We aim to explore the performance of of the 25 most popular measures in the graph clustering problem on a set of generated graphs with different parameters. We assess the quality of clustering with every considered measure and determine the best measure for every region of the graph parameter space.

Our contributions are as follows:
\begin{itemize}
    \item We generate a dataset of \camready{11780} graphs covering \camready{the entire} parameter space of the LFR generator;
    \item We consider a broad set of measures and rank them by clustering performance on this dataset;
    \item We determine the graph features that are responsible for measure leadership;
    \item We find the regions of each measure's leadership in the graph parameter space.
\end{itemize}

Our framework for clustering with graph measures as well as the collected dataset are available \camready{at}~\url{https://github.com/vlivashkin/pygkernels}.

\section{Definitions}

\subsection{Kernel $k$-means}
The original $k$-means algorithm~\cite{lloyd1982least,macqueen1967some} clusters objects in Euclidean space. It requires coordinates of the objects to determine the distances between them and centroids.
The algorithm can be generalized to use the degree of closeness between the objects without defining a particular space. This technique is called \textit{the kernel trick}, usually it is used to bring non-linearity to linear algorithms. The algorithm that uses the kernel trick is called \textit{kernel $k$-means} (see, e.g., \cite{fouss2016algorithms}). For graph node clustering scenario, we can use graph measures as kernels for the kernel $k$-means.

Initially, the number of clusters is known and we need to set the initial state of centroids.
The results of clustering with $k$-means are very sensitive to the initial state. Usually, the algorithm runs several times with different initial states (trials) and chooses the ``best'' trial. There are different approaches to the initialization; we consider three of them: random data points, $k$-means++~\cite{arthur2006k}, and random partition.
We combine all these strategies to reduce the impact of the initialization strategy on the result.

\subsection{Closeness measures}
\lblsec{measures}
For a given graph $G$, $V(G)$ is the set of its vertices and $A$ is its adjacency matrix.
A \textit{measure} on $G$ is a function $\kappa\!: V(G)\times V(G)\to\R$, which gets two nodes and returns closeness (\camready{larger} means closer) or distance (\camready{larger} means farther).

A \textit{kernel on a graph} is a graph nodes' closeness measure that has an inner product representation. Any symmetric positive semidefinite matrix is an inner product matrix (also called Gram matrix). A kernel matrix $K$ is a square matrix that contains similarities for all pairs of nodes in a graph.

To use kernel $k$-means, we need kernels. Despite that not all closeness measures we consider are Gram matrices, we treat them as kernels. The applicability of this approach was confirmed in~\cite{fouss2016algorithms}. For the list of measures \camready{below}, we use the word ``kernel'' only for the measures that satisfy the precise definition of kernel.

\camready{Classical measures \textit{Shortest Path} distance~\cite{buckley1990distance} (SP) and \textit{Commute Time} distance~\cite{gobel1974random} (CT) are} defined as distances, so we need to transform them into similarities to use as kernels. We apply the following distance to closeness transformation~\cite{chebotarev2005duality,borg2005modern}:
\begin{equation}
    K = -H\mathcal{D}H;\; H = I - E/n,
    \label{eq:dist2kernel}
\end{equation}
where $\mathcal{D}$ is a distance matrix, $E$ is the matrix of ones, $I$ is the identity matrix, and $n$ is the number of nodes.

In this paper, we examine 25 graph measures (or, more exactly, 25 parametric families of measures). We present these measures grouped by type similarly to~\cite{avrachenkov2017kernels}:
\begin{itemize}
    \item Adjacency Matrix $A$ based kernels and measures.
        \begin{itemize}
            \item \textit{Katz kernel}:
                $K_\alpha^{\text{Katz}} = (I - \alpha A)^{-1}$, $0 < \alpha < \rho^{-1}$, where
                $\rho$ is the spectral radius of $A$~\cite{katz1953new} (also known as Walk proximity~\cite{chebotarev2006proximity} or von Neumann diffusion kernel~\cite{kandola2003learning,shawe2004kernel}).
            \item \textit{Communicability kernel}
                $K_t^{\text{Comm}} = \expm(t A)$, $t>0$, where
                $\expm$ means matrix exponential                ~\cite{fouss2006experimental,estrada2007statistical,estrada2008communicability}.
            \item \textit{Double Factorial closeness}:
                $K_t^{\text{DF}} = \sum_{k=0}^{\camready{\infty}}\frac{t^k}{k!!} A^k$, $t>0$ \cite{estrada2017accounting}.
        \end{itemize}
    \item Laplacian Matrix $L = D - A$ based kernels and measures, \camready{where} $D=\Diag(A\cdot\b{1})$ is the degree matrix of $G$, $\Diag(\b{x})$ is the diagonal matrix with vector $\b{x}$ on the main diagonal.
        \begin{itemize}
            \item \textit{Forest kernel}:
                $K_t^{\text{For}} = (I + t L)^{-1}$, $t>0$ (also known as Regularized Laplacian kernel)~\cite{chebotarev1995proximity}.
            \item \textit{Heat kernel}:
                $K_t^{\text{Heat}} = \expm(-t L)$, $t>0$~\cite{chung1998coverings}.
            \item \textit{Normalized Heat kernel}:
                $K_t^{\text{NHeat}} = \expm(-t \mathcal{L})$,
                $\mathcal{L} = D^{-\frac12} L D^{-\frac12}$,\\ $t>0$~\cite{chung1997spectral}.
            \item \textit{Absorption kernel}:
                $K_t^{\text{Abs}} = (t A + L)^{-1}$, $t>0$~\cite{jacobsen2018generalized}.
        \end{itemize}
    \item Markov Matrix $P = D^{-1} A$ based kernels and measures.
        \begin{itemize}
            \item \textit{Personalized PageRank closeness}:
                $K_\alpha^{\text{PPR}} = (I-\alpha P)^{-1}$,\\
                $0<\alpha<1$~\cite{page1999pagerank}.
            \item \textit{Modified Personalized PageRank}:
                $K_\alpha^{\text{MPPR}} = (I-\alpha P)^{-1}D^{-1} = (D-\alpha A)^{-1}$, 
                $0<\alpha<1$ \cite{kirkland2012group}.
            \item \textit{PageRank heat closeness}:
                $K_t^{\text{HPR}} = \expm(-t(I-P))$,
                $t>0$ \cite{chung2007heat}.
            \item \textit{Randomized Shortest Path distance}. Using $P$ and the matrix of the SP distances $C$ first get $Z$~\cite{yen2008family}:
                \begin{equation}
                    \camready{
                    Z = (I - W)^{-1}
                    \text{, where }
        			W = P \circ \exp(-\beta C)\text{.}
        			}
        			\label{eq:rsp_base}
        		\end{equation}
        		Then
        		$S = (Z(C \circ W)Z) \div Z$;
        		$\bar{C} = S - \b{e}\diag(S)^T$, and finally,
        		$\mathcal{D}_\text{RSP} = (\bar{C} + \bar{C}^T)/2$.
        		Here $\circ$ and $\div$ are element-wise multiplication and division, \camready{ $\diag(S)$ is the column vector of the diagonal elements of $S$.}
        		The kernel version $K^{\text{RSP}}(t)$ can be obtained with \eqref{eq:dist2kernel}.
            \item \textit{Free Energy distance}. Using $Z$ from \eqref{eq:rsp_base}:
                $\widetilde{Z} = Z\Diag(Z)^{-1}$;
        		$\Phi = -1/\beta \log{\widetilde{Z}}$;
        		$\mathcal{D}_\text{FE} = (\Phi + \Phi^T)/2$~\cite{kivimaki2014developments}.
        		Kernel version $K^{\text{FE}}(t)$ can be obtained with \eqref{eq:dist2kernel}.
        \end{itemize}
    \item Sigmoid Commute Time kernels.
        \begin{itemize}
            \item \textit{Sigmoid Commute Time kernel}:
                \begin{equation}
                    K_t^{\text{SCT}} = \sigma(-t K^{\text{CT}} / \text{std}(K^{\text{CT}})),\:
                    t > 0,
                    \label{eq:sct}
                \end{equation}
                where $K^{\text{CT}}$ is the matrix of Commute Time proximity, std is the standard deviation of all elements of a matrix\camready{,} $\sigma$ is the element-wise sigmoid function $\sigma(x) = 1/(1+e^{-x})$~\cite{yen2007graph}.
            \item \textit{Sigmoid Corrected Commute Time kernel}.
                First of all, we need the Corrected Commute Time kernel~\cite{luxburg2010getting}:
             	$K^\text{CCT} = H D^{-\frac12} M(I-M)^{-1} M D^{-\frac12} H$, where
             	$M = D^{-\frac12} \Big(A - \frac{\vec{\b{d}}\vec{\b{d}}^T}{\vol(G)}\Big) D^{-\frac12}$,
                $H$ is the centering matrix $H = I - E/n$, $\vec{\b{d}}$ is the vector of diagonal elements of $D$ and $\vol(G)$ is the sum of all elements of $A$.
                Then, apply \eqref{eq:sct} replacing $K^{\text{CT}}$ with $K^{\text{CCT}}$ to obtain $K^{\text{SCCT}}$.
        \end{itemize}
\end{itemize}

Occasionally, element-wise logarithm is applied to the resulting kernel matrix~\cite{chebotarev2013studying,ivashkin2016logarithmic}. We apply it to almost all investigated measures and consider the resulting measures separately from their plain versions~(see \reftbl{kernels}). For some measures, like Forest kernel, this is well-known practice~\cite{chebotarev2013studying}, while for others, like Double Factorial closeness, this transformation, to the best of our knowledge, is applied for the first time. The considered measures and their short names are summarized in~\reftbl{kernels}.

\begin{table}
    \caption{Short names of considered kernels and other measures. }
    \lbltbl{kernels}
    \centering
    \begin{tabular}{l|l|l|p{6.0cm}}
         & \multicolumn{2}{c|}{Short name} & \\
        Family & Plain & Log & Full name \\
    	\hline
    	\multirow{2}{2.5cm}{Adjacency matrix based kernels}
    	  & Katz & logKatz & Katz kernel \\
    	  & Comm & logComm & Communicability kernel \\
    	  & DF & logDF & Double Factorial closeness \\
    	\hline
    	\multirow{3}{2.5cm}{Laplacian based kernels}
    	  & For & logFor & Forest kernel \\
    	  & Heat & logHeat & Heat kernel \\
    	  & NHeat & logNHeat & Normalized Heat kernel \\
    	  & Abs & logAbs & Absorption kernel \\
    	\hline
    	\multirow{3}{2.5cm}{Markov matrix based kernels and measures}
    	  & PPR & logPPR & Personalized PageRank closeness\\
    	  & MPPR & logMPPR & Modified Personalized PageRank \\
    	  & HPR & logHPR & PageRank heat closeness \\
      	  & RSP & - & Randomized Shortest Path kernel \\
    	  & FE & - & Free Energy kernel \\
    	\hline
    	\multirow{2}{2.5cm}{Sigmoid Commute Time}
    	  & SCT & - & Sigmoid Commute Time kernel \\
    	  & SCCT & - & Sigmoid Corrected Commute Time kernel \\
    	\hline
    	& SP-CT & - & Linear combination of SP and CT \\
    \end{tabular}
\end{table}

\section{Dataset}
We collected a paired dataset of graphs and the corresponding results of clustering with each measure mentioned in \reftbl{kernels}. In this section, we describe the graph generator, the sampling strategy, the graph characteristic features, and the pipeline for the measure score calculation.

We use Lancichinetti--Fortunato--Radicchi (LFR) graph generator. It generates non-weighted graphs with ground truth non-overlapping communities. The model has five mandatory parameters: the number of nodes $n$ ($n > 0$), the power law exponent for the degree distribution $\tau_1$ ($\tau_1 > 1$), the power law exponent for the community size distribution $\tau_2$ ($\tau_2 > 1$), the fraction of intra-community edges incident to each node $\mu$ ($0 \leq \mu \leq 1$), and either minimum degree (min degree) or average degree (avg degree). There are also several extra parameters: maximum degree (max degree), minimum community size (min community), maximum community size (max community). Not the whole LFR parameter space corresponds to common real-world graphs; most of such graphs are described with $\tau_1 \in [1, 4]$ and $\mu < 0.5$~(e.g.,~\cite{fotouhi2019evolution}). However, there is also an interesting case of bipartite/multipartite-like graphs with $\mu > 0.5$. Our choice is to consider the entire parameter space to cover all theoretical and practical cases.

For the generation, we consider $10 < n < 1500$. It is impossible to generate a dataset with a uniform distribution of all LFR parameters, because $\tau_1$ and $\tau_2$ parameters are located on rays. We transform $\tau_1$ and $\tau_2$ to \camready{$\tilde{\tau_i} = 1 - (1 / \tau_i^{0.7}),i=1,2$} to bring their scope to the $[0, 1]$ interval. In this case, ``realistic'' settings with $\tau_1 \in [1, 4]$ take up \camready{62\%} of the variable range. Also, as avg degree feature is limited by the number of nodes $n$ of a particular graph, we decided to replace it with density ($\text{avg degree}/(n-1)$). It belongs to $[0, 1]$. Using all these considerations, we collected our dataset by uniformly sampling parameters for LFR generator from the set \camready{\{}$n$, $\tilde{\tau}_1$, $\tilde{\tau}_2$, $\mu$, density\camready{\}} and generating graphs with these parameters. Additionally, we filter out all disconnected graphs.

In total, we generated \camready{11780} graphs. It is worth noting that the generator fails for some sets of parameters, so the resulting dataset is not uniform~(see \reffig{dataset_lfr_features_distribution}). In our study, non-uniformity is not a \camready{critical} issue, because we are interested in local effects, rather than global leadership. Moreover, true uniformity for LFR parameter space is impossible, due to the unlimited scope of parameters.

\begin{figure}[h]
    \centering
    \begin{minipage}[b]{0.44\textwidth}
        \includegraphics[width=\linewidth]{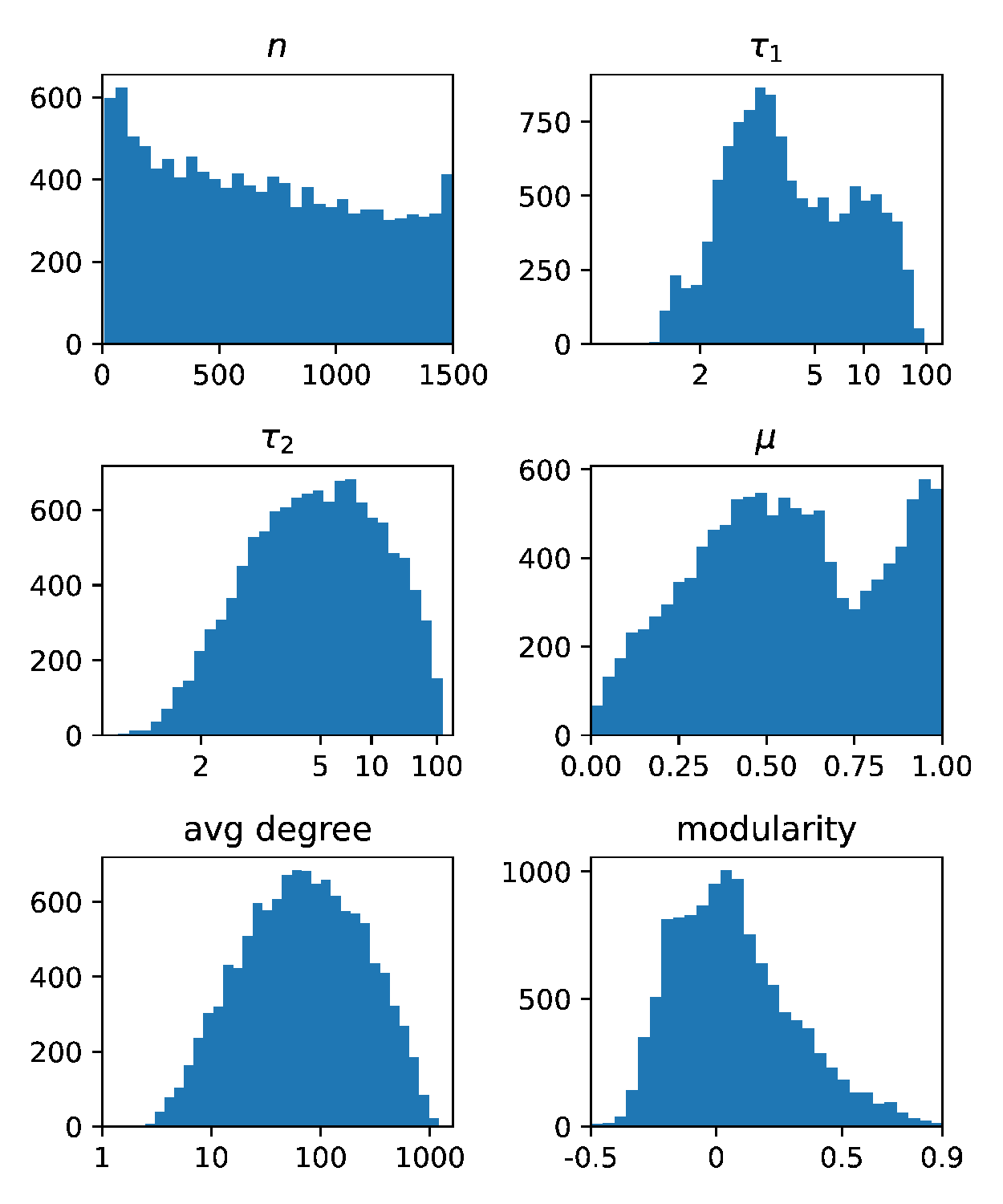}
        \caption{Distribution of graph features in the dataset\\}
        \lblfig{dataset_lfr_features_distribution}
    \end{minipage} \hspace{.01\linewidth} %
    \begin{minipage}[b]{0.52\textwidth}
    	\resizebox{\textwidth}{!}{
    	\begin{tikzpicture}
            \begin{scope}[every node/.style={rectangle,thick,draw}]
                \node (lfr_params) at (2,4.8) {\begin{tabular}{c}
                     LFR parameters \\
                     $n$, $tau1$, $tau2$, $mu$, etc.
                \end{tabular}};
                \node (lfr) at (2,3.5) {LFR};
                \node (graph) at (2,2.2) {\begin{tabular}{c}
                graph \& ground \\ truth partition
                \end{tabular}};
                \node (measure_param) at (-2.2,3.8) {\begin{tabular}{c}
                    measure parameter \\
                    $t \in [0,1]$
                \end{tabular}};
                \node (measure) at (-2.2,2.2) {\begin{tabular}{c}
                    measure \\
                    (Katz, Forest etc.)
                \end{tabular}};
                \node (init) at (-3.8,0) {\begin{tabular}{c}
                    initialization strategy \\
                    ($k$-means++, etc.)
                \end{tabular}};
                \node (kmeans) at (0,0) {\begin{tabular}{c}
                    \large\b{k-means} \\
                    (several trials)
                \end{tabular}};
                \node (criteria) at (0,-1.7) {\begin{tabular}{c}
                    trial choosing criteria \\
                    (inertia, modularity)
                \end{tabular}};
                \node (score) at (2,-3.6) {ARI score};
            \end{scope}
            \begin{scope}[>={Stealth[black]},
                          every node/.style={fill=white,rectangle},
                          every edge/.style={draw=red,very thick}]
                \path [->] (lfr_params) edge (lfr);
                \path [->] (lfr) edge (graph);
                \path [->] (graph) edge node {$A$} (measure);
                \path [->] (graph) edge node {$k_{\rm clusters}$} (kmeans);
                \path [->] (graph) edge node {$y_{\rm true}$} (score);
                \path [->] (measure_param) edge (measure);
                \path [->] (measure) edge node {$K$} (kmeans);
                \path [->] (init) edge (kmeans);
                \path [->] (kmeans) edge (criteria);
                \path [->] (criteria) edge node {$y_{\rm pred}$} (score);
            \end{scope}
        \end{tikzpicture}
        }
        \caption{Measuring ARI clustering score for a particular graph, measure, and measure parameter}
        \lblfig{bench}
    \end{minipage}
\end{figure}

For our research, we choose a minimum set of the features that describe particular properties of graphs and are not interchangeable.

The LFR parameters can be divided in\camready{to} three groups by the graph properties they reflect:
\begin{itemize}
    \item The size of the graph and the communities: $n$, $\tau_1$, min community, max community;
    \item The density and uniformity of the node \camready{degree} distribution: $\tau_2$, min degree, avg degree, max degree. As avg degree depends on $n$, it is distributed exponentially, so we use $\log(\text{avg degree})$ instead;
    \item The cluster separability: $\mu$. As $\mu$ parameter considers only the ratio between the number of inter-cluster edges and the number of nodes but ignores overall density, we use modularity~\cite{newman2004finding} as a more appropriate measure for cluster separability.
\end{itemize}

Thus, the defined set of features \camready{\{}$n$, $\tau_1$, $\tau_2$, avg degree, modularity\camready{\}} is enough to consider all graph properties mentioned above. Although modularity is a widely used measure, it suffers from resolution limit problems~\cite{fortunato2007resolution}. We acknowledge that this may cause some limitations in our approach, which should be the topic of further research.

For every generated graph, we calculate the top ARI score for every measure~\cite{hubert1985comparing}. The popular NMI clustering score is known to be biased towards smaller clusters, according to~\cite{gosgens2019systematic}. We choose ARI as a clustering score \camready{that} is both well known and unbiased. As soon as every measure has a parameter, we perform clustering for a range of parameter values (we transform the parameter to become in the [0, 1] interval and then choose 16 values linearly spaced from 0 to 1). For each value, we run $6+6+6$ trials of $k$-means (6 trials for each of three initialization methods).

\reffig{bench} shows the pipeline we use to calculate ARI score for a given LFR parameter set, a measure, and a measure parameter. Measure parameters are not the subject of our experiments, so for every measure we just take the result of the measure with the value of the parameter that gives the best ARI score.

Because of the need to iterate over graphs, measures, parameter values, and initializations, the task is quite computationally complex. The total computation time was 20 days on 18 CPU cores and 6 GPUs.

\section{Results}

\subsection{Global leadership in LFR space}
As a rough estimate of the measures' applicability, we calculate global leadership on our generated dataset. We divide the dataset into parts corresponding to two different cases of clustering: associative (graphs with modularity $\geq 0$ on the ground truth partition) and the \camready{complementary} dissociative case. \camready{The negative modularity case basically corresponds to the $\mu > 0.5$ setup of LFR generator.}

We \camready{rank} the measures by their ARI score on every graph of the dataset. The aggregated rank is defined as the position of the measure in this list, averaged over the dataset (see Tables~\ref{tbl:lb_associative} and \ref{tbl:lb_dissociative}; smaller rank is better). It is important to note that the global leadership does not give a comprehensive advice on which measure is better to use, because for a particular graph, the global leader can perform worse than the \camready{local winner}. Here, we consider the entire LFR space, not just its zone corresponding to common real-world graphs, so the ranking may differ from those obtained for restricted settings.

\begin{table}[h]
    \caption{Leaderboards for associative and dissociative cases. The ARI column shows the mean ARI across the dataset.}

    \centering
    \begin{subtable}{.46\linewidth}
        \begin{tabular}{rlrrr}
            \hline
            \# & Measure & Rank & Wins, \% & ARI \\
            \hline
             1 & RSP & 4.1 & 40.0 & 0.67 \\
             2 & SCCT & 5.1 & 50.5 & 0.68 \\
             3 & logNHeat & 5.3 & 34.7 & 0.66 \\
             4 & logHeatPR & 5.3 & 34.9 & 0.66 \\
             5 & FE & 5.5 & 35.8 & 0.66 \\
             6 & logKatz & 5.5 & 39.5 & 0.66 \\
             7 & logPPR & 6.2 & 35.1 & 0.65 \\
             8 & logComm & 6.3 & 40.5 & 0.64 \\
             9 & logModifPPR & 6.5 & 34.4 & 0.65 \\
            10 & SCT & 7.3 & 36.1 & 0.64 \\
            11 & SP-CT & 7.5 & 32.6 & 0.64 \\
            12 & logAbs & 8.1 & 33.8 & 0.63 \\
            13 & logFor & 8.8 & 33.8 & 0.60 \\
            14 & logHeat & 9.2 & 31.1 & 0.58 \\
            15 & NHeat & 9.6 & 35.1 & 0.56 \\
            16 & HeatPR & 10.3 & 32.2 & 0.59 \\
            17 & Comm & 11.5 & 26.4 & 0.52 \\
            18 & logDF & 12.4 & 22.7 & 0.46 \\
            19 & Heat & 13.7 & 27.2 & 0.46 \\
            20 & Katz & 15.2 & 10.1 & 0.43 \\
            21 & DF & 16.0 & 12.1 & 0.37 \\
            22 & PPR & 17.8 & 11.1 & 0.35 \\
            23 & For & 20.4 & 7.9 & 0.19 \\
            24 & Abs & 21.1 & 7.1 & 0.16 \\
            25 & ModifPPR & 22.1 & 4.7 & 0.12 \\
            \hline
        \end{tabular}
        \caption{Associative graphs. The win percentage is calculated among 6777 graphs in the dataset.}
        \lbltbl{lb_associative}
    \end{subtable} \hspace{.02\linewidth} %
    \begin{subtable}{.46\linewidth}
        \centering
        \begin{tabular}{rlrrr}
            \hline
            \# & Measure & Rank & Wins, \% & ARI \\
            \hline
             1 & SCCT & 3.7 & 63.9 & 0.70 \\
             2 & RSP & 7.0 & 18.6 & 0.46 \\
             3 & SP-CT & 8.1 & 16.6 & 0.45 \\
             4 & SCT & 8.1 & 14.5 & 0.46 \\
             5 & NHeat & 8.5 & 10.1 & 0.40 \\
             6 & logHeatPR & 8.6 & 14.5 & 0.41 \\
             7 & FE & 8.9 & 15.5 & 0.43 \\
             8 & logNHeat & 9.0 & 13.9 & 0.39 \\
             9 & logPPR & 9.4 & 13.8 & 0.39 \\
            10 & Katz & 9.8 & 2.8 & 0.34 \\
            11 & Comm & 10.3 & 5.2 & 0.33 \\
            12 & logModifPPR & 10.8 & 13.5 & 0.37 \\
            13 & logKatz & 10.9 & 14.4 & 0.37 \\
            14 & Abs & 11.0 & 13.3 & 0.35 \\
            15 & DF & 12.2 & 2.6 & 0.27 \\
            16 & logAbs & 12.8 & 12.1 & 0.35 \\
            17 & HeatPR & 15.9 & 0.6 & 0.16 \\
            18 & logFor & 15.9 & 5.3 & 0.22 \\
            19 & Heat & 15.9 & 0.7 & 0.09 \\
            20 & PPR & 16.4 & 0.5 & 0.15 \\
            21 & logHeat & 16.6 & 0.7 & 0.10 \\
            22 & logDF & 18.1 & 3.7 & 0.08 \\
            23 & logComm & 18.2 & 0.7 & 0.07 \\
            24 & For & 21.4 & 0.0 & 0.02 \\
            25 & ModifPPR & 21.7 & 0.1 & 0.02 \\
            \hline
        \end{tabular}
        \caption{Dissociative graphs. The win percentage is calculated among 5003 graphs in the dataset.}
        \lbltbl{lb_dissociative}
    \end{subtable}
\end{table}

\reftbl{lb_associative} shows that there are several leading measures whose quality is not much different. The best measures are RSP (by rank) and SCCT (by the number of wins and the mean ARI).
Dissociative case has an undisputed leader, SCCT \camready{(\reftbl{lb_dissociative})}. 

The above \camready{division} into two cases does not exhaust all the variety of graphs. For the further more precise study of the measures' applicability, we will look for the leadership zones of each measure.

\subsection{Feature importance study}
First of all, we find out which graph features among the LFR parameters are important for the choice of the best measure and which are not. To do that, we use Linear Discriminant Analysis~\cite{mika1999fisher} (LDA). This method finds a new basis in the feature space to classify a dataset in the best way. It also shows how many components of basis are required to fit the majority of data.

\begin{figure}[h]
    \centering
    \begin{subfigure}[b]{0.32\textwidth}
        \centering
        \includegraphics[width=\columnwidth]{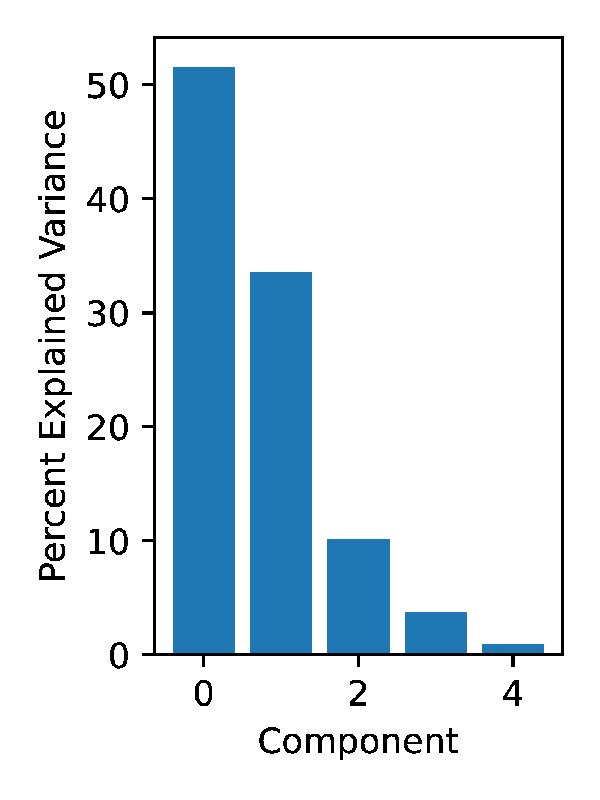}
        \caption{Explained variance.}
        \lblfig{lda_explained_variance}
    \end{subfigure} %
    \begin{subfigure}[b]{0.55\textwidth}
        \centering
        \includegraphics[width=\columnwidth]{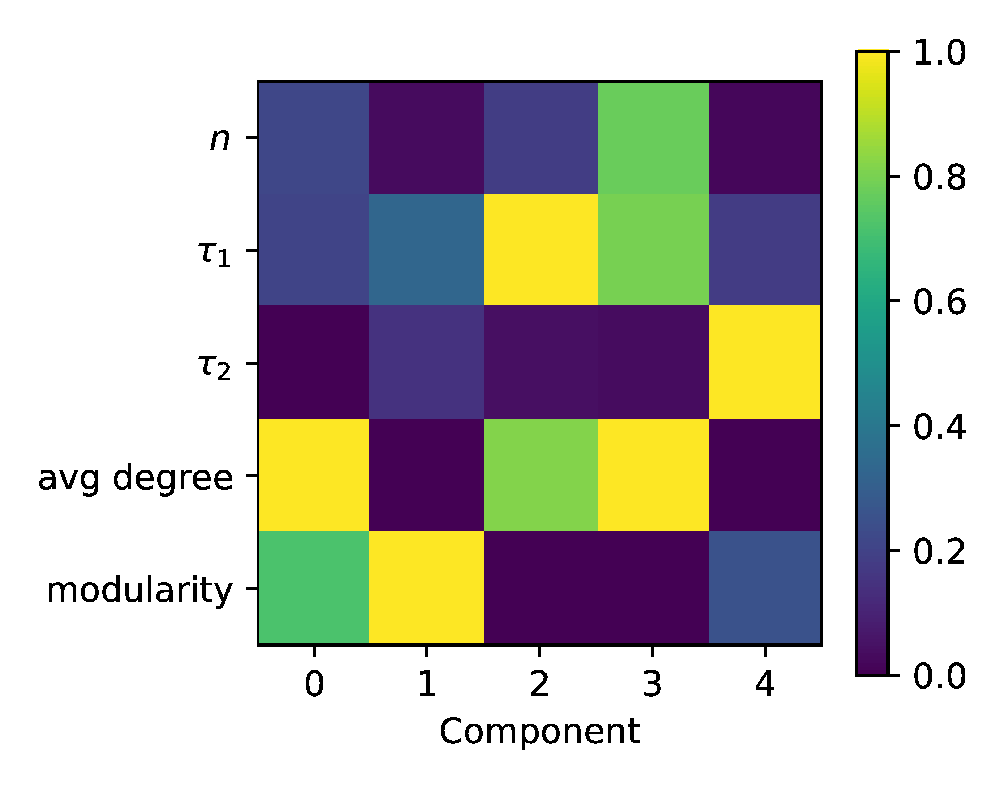}
        \caption{Features' contribution to LDA components.}
        \lblfig{lda_feature_contribution}
    \end{subfigure}
    \caption{The results of LDA \camready{analysis.}}
    \lblfig{lda_feature_importance}
\end{figure}

\reffig{lda_explained_variance} shows that the first two components \camready{account for} about 90\% of the explained variance. \reffig{lda_feature_contribution} shows that these components include only $\tau_1$, avg degree, and modularity. The fact that $n$ is not included means that the size of the graph as well as the density are not of primary importance for choosing the best measure. 
So is not $\tau_2$ measuring the diversity of cluster sizes.

To detect the zones of measure leadership, we need to know the leadership on average in each area of space rather than the wins in particular points. To determine the local measure leadership, we need to introduce a filtering algorithm that for every point of the space returns the leading measure depending on the closest data points. As the choice of measure is mainly dependent on three features \camready{\{}$\tau_1$, avg degree, modularity\camready{\}}, we can limit our feature space to them.

\subsection{Gaussian filter in feature space}
Using a filter in the feature space, we can suppress the noise and reveal the actual zones of leadership for the measures. We use the Gaussian filter with a scale parameter $\sigma$. For every given point of the space, it takes the data points that are closer than $3\sigma$ and averages ARIs of the chosen points with a weight $e^{-\text{dist}^2/2\sigma^2}$. This allows to give larger weights to closer points. If there are less than three data points inside the sphere with a $3\sigma$ radius, the filter returns nothing, allowing to ignore the points with insufficient data in their vicinity.

Before applying the filter, we prepare the dataset. First, we isolate the case when several measures reach $\text{ARI}=1$ into a separate measure called ``several''. Also, we normalize the standard deviation of every feature distribution to one.

To choose $\sigma$, we apply the filter with different \camready{values of} $\sigma$ and look at the number of connected components in the feature space.
The needed $\sigma$ should be large enough to suppress the noise, however, it should not suppress small zones. Guided by this heuristic, we choose $\sigma = 0.6$.

\begin{table}[h]
    \caption{The leaderboard of measure wins after filtering with $\sigma=0.6$}
    \lbltbl{lbs_after_filtering}
    \centering
    \begin{tabular}{l|rrrrrrrrrrr}
     & \rotatev{SCCT} & \rotatev{RSP} & \rotatev{logComm} & \rotatev{logHeatPR} &  \rotatev{several} & \rotatev{Abs} & \rotatev{logHeat} & \rotatev{logNHeat} & \rotatev{NHeat}  & \rotatev{Comm} \\
    \hline
    Wins & 7874 & 1544 & 1043 & 444 & 265 & 10 & 4 & 3 & 2 & 2 \\
    \end{tabular}
\end{table}

After filtering with $\sigma=0.6$, the leaderboard of measure wins changes (see \reftbl{lbs_after_filtering}). Only \camready{four} measures keep their positions: SCCT, RSP, logComm, and logHeatPR. \camready{There is also a special case of several winning measures (named ``several''), when the predicted partition reaches $\text{ARI}=1$ for several measures. The presence of several winners makes it difficult to analyze zones of measure's leadership, so we decided to exclude this case from the detailed analysis in this work. Filtering shows that} these \camready{four} measures do have zones of leadership, otherwise they would be filtered out. We can plot the entire feature space colored by the leadership zones of the measures~(see \reffig{splits}). As the resulting space is 3D, we indicate its slices by their coordinates.

\begin{figure*}[h]
    \centering
    \begin{subfigure}[b]{1\textwidth}
        \centering
        \includegraphics[width=0.88\columnwidth]{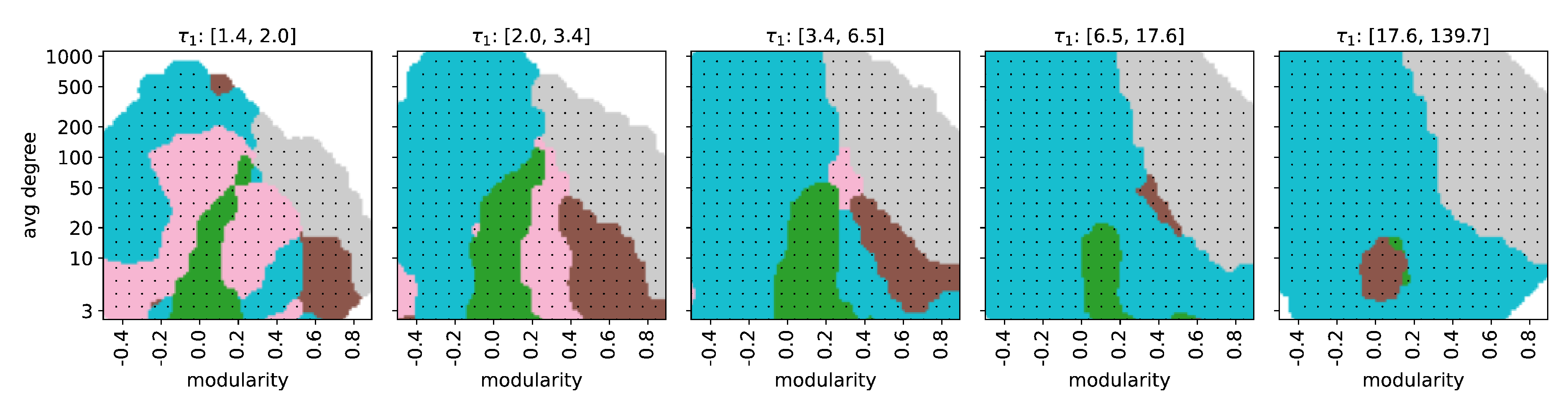}
        \includegraphics[width=0.11\columnwidth]{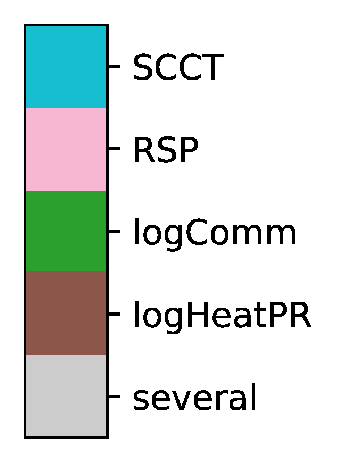}
        \caption{Slices by $\tau_1$.}
        \lblfig{splits_by_tau1}
    \end{subfigure}\\
    \begin{subfigure}[b]{1\textwidth}
        \centering
        \includegraphics[width=0.88\columnwidth]{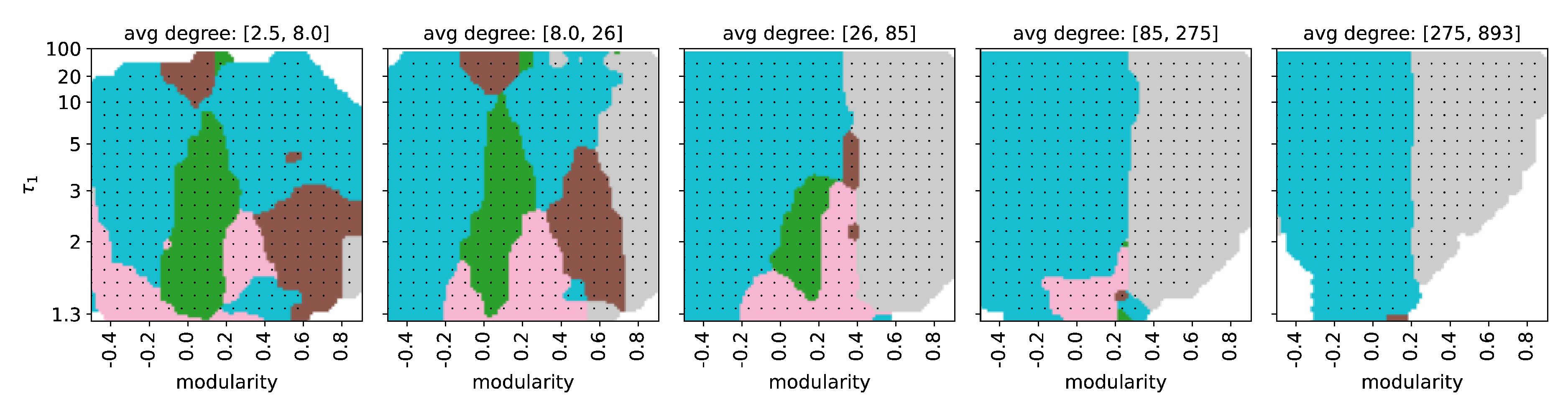}
        \includegraphics[width=0.11\columnwidth]{legend.pdf.eps}
        \caption{Slices by avg degree.}
        \lblfig{splits_by_avg_degree}
    \end{subfigure}\\
    \begin{subfigure}[b]{1\textwidth}
        \centering
        \includegraphics[width=0.88\columnwidth]{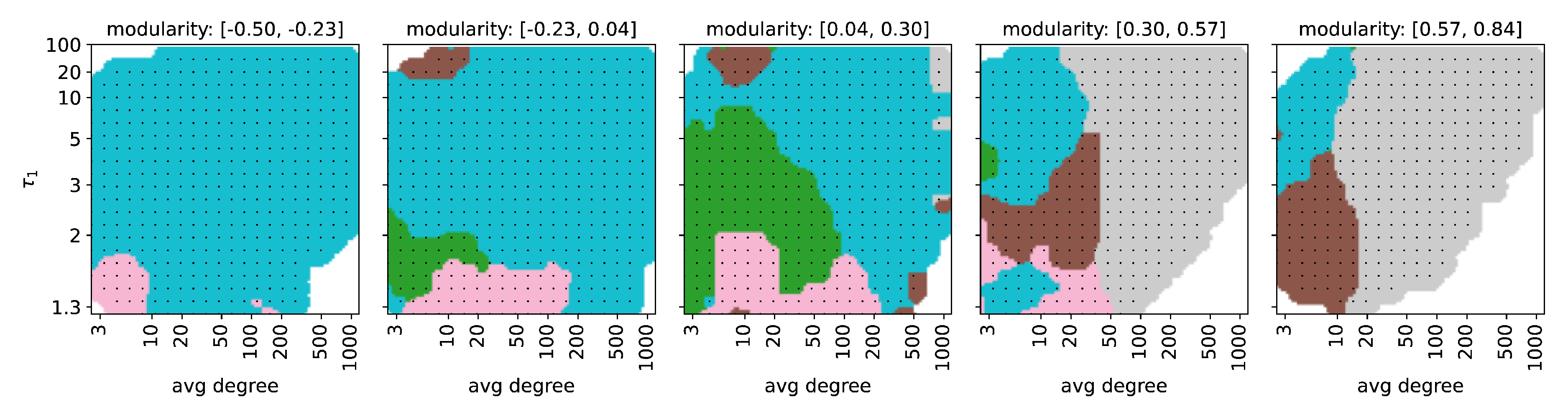}
        \includegraphics[width=0.11\columnwidth]{legend.pdf.eps}
        \caption{Slices by modularity.}
        \lblfig{splits_by_modularity}
    \end{subfigure}
    \caption{The feature space \camready{\{}$\tau_1$, avg degree, modularity\camready{\}} divided into the leadership zones of six measures.}
    \lblfig{splits}
\end{figure*}

The zones of measure leadership can be described by the following approximate criteria:
\begin{itemize}
    \item SCCT: \camready{in many domains of the parameter space};
    \item RSP: $\tau_1$ up to 5, modularity in $-0.25..0.3$;
    \item logComm: modularity in 0..0.3, avg degree up to 100;
    \item logHeatPR: modularity above 0.3, avg degree up to 50;
    \item several: \camready{high} modularity or \camready{high} avg degree.
\end{itemize}

\section{Conclusions}
In this work, we have shown that the global leadership of measures does not provide comprehensive knowledge about graph measure performance \camready{in clustering tasks}. We demonstrated that among 25 measures, SCCT is the best measure for the LFR graphs both by winning rate and ranking. However, there are also smaller \camready{distinct} zones of leadership for RSP, logComm, and logHeatPR. \camready{Other measures, including those with high rank, fail to form their leadership zones.}

Our results do not contradict those of other experimental works and\camready{, moreover, } refine them by providing new findings. LogComm was first introduced in~\cite{ivashkin2016logarithmic} and won in the competitions on graphs generated with a particular set of SBM parameters. This study confirms its leadership, but only for a certain type of graphs. Another interesting finding is logHeatPR, which shows unexpectedly good performance \camready{within} its zone of leadership.

\camready{Accoring to LDA analysis results, the leadership of measure is determined mainly by \{$\tau_1$, avg degree, modularity\}. One of the interesting consequences is that the leadership does not depend on $n$. This effect could be caused by the fact that we limited the size of the graphs to $n < 1500$. It is not guaranteed to be preserved for large graphs.}

This study is based on the LFR benchmark data. \camready{More research is needed to determine how well the results for LFR fit with to real-world. This would assess the applicability of our findings to practical cases.}

It should be noted that our study is insensitive to the non-uniformity of the generated dataset. While manipulations with this dataset may affect the global leaderboard, they cannot change the local leadership, which is the focus of the present work.

\bibliographystyle{splncs03} 
\bibliography{references}

\end{document}